\documentstyle[12pt]{article}

\begin{document}

\title{TIME-SYMMETRIZED COUNTERFACTUALS IN QUANTUM THEORY}
\author{ Lev Vaidman}
\date{}
\maketitle

\begin{center}
{\small \em School of Physics and Astronomy \\
Raymond and Beverly Sackler Faculty of Exact Sciences \\
Tel Aviv University, Tel-Aviv 69978, Israel. \\}
\end{center}

\vspace{2cm}
\begin{abstract}
Counterfactuals in quantum theory are briefly reviewed and it is
argued that they are very different from counterfactuals considered in
the general philosophical literature. The issue of time symmetry of quantum
counterfactuals is considered  and a novel time-symmetric definition of
quantum counterfactuals is proposed. This definition is applied for
analyzing  several controversies related to quantum counterfactuals.
\end{abstract}

\vfill\break

There are very many philosophical discussions on the concept of
counterfactuals and especially on the time's arrow in counterfactuals.
There is also a considerable literature on counterfactual in quantum
theory.  In order to be a helpful tool in quantum theory
counterfactuals have to be rigorously defined. Unfortunately, the
concept of counterfactuals is vague\footnote{``Counterfactuals are
  infected with vagueness, as everybody agrees.''(Lewis, 1986, 34)}
and this leads to several controversies.  I, however, believe that
since quantum counterfactuals appear in a much narrow context than
in general discussions on counterfactuals, they can be defined
unambiguously.  I will briefly review counterfactuals in quantum
theory and will propose a rigorous definition which  can
clarify several issues, in particular, those related to the
time-symmetry of quantum counterfactuals.

 A general form of a counterfactual
is
\begin{quotation}
{ \bf (i)}
{\em If it were that $\cal A$, then it would be that $\cal B$.}
\end{quotation}
\noindent
The basic approach to analyzing counterfactuals is to consider the
{\it actual} world, the world that  we know,  in which $\cal A$ is in general
not true, and a counterfactual world, {\it closest} to the actual
world, in which $\cal A$ is true. The truth of the counterfactual (i)
depends on the truth of $\cal B$ in this counterfactual world.

There is  a general
philosophical trend to consider counterfactuals to be asymmetric in
time. Even Bennett who was challenging this claim in 1984 reversed his
position (as I learned from private correspondence). In the
most influential paper on this subject, Lewis  writes:
\begin{quotation}
I believe that indeterminism is neither necessary nor sufficient for
the asymmetries I am discussing. Therefore I shall ignore the
possibility of indeterminism in the rest of this paper, and see how
the asymmetries might arise even under strict determinism. (1986, 37) 
\end{quotation}
In contrast to this opinion, I  believe that the indeterminism is crucial for allowing
nontrivial time-symmetric counterfactuals, and that Lewis's and other
general philosophical analyses are irrelevant for the issue of
counterfactuals in quantum theory.  The key questions in these
analyses are related to $\cal A$: How come $\cal A$ if in the actual world
$\cal A$ is not true?  Do we need a ``miracle'' (i.e. breaking the
laws of
physics) for $\cal A$?  Does $\cal A$ come by itself, or it is
accompanied by other changes?  In contrast, in the context of quantum
theory there are no important questions related to $\cal A$. In some
cases $\cal A$ is related to an external entity which might vary freely by
fiat, in other cases the indeterminism of the theory allows different
$\cal A$ without need for ``miracles'' -- the main topic of discussion
on counterfactuals in general philosophy.

The main source of vagueness in counterfactuals is in the definition
of a counterfactual world {\it closest} to the actual world. Clearly,
it differs in $\cal A$. In a
deterministic world  other differences are also required: a
``miracle'' for $\cal A$ to happen, etc. There is no 
rigorous specification of   aspects of  a counterfactual world
which are fixed to
be identical to those of the actual world. The definition of
such specification is
missing in most discussions on quantum counterfactuals too. The main result
of this work is a proposal for such definition. The most important
feature of this definition  is that it is also applicable for
time-symmetric situations.

In the literature on quantum theory there are two main (different)
concepts named ``counterfactuals''.  Quantum counterfactuals of the
first type are events which did not happen in our world, but somehow
influenced it. To present this concept let me quote Penrose:
\begin{quotation}
  \noindent
What is particularly curious about quantum theory is that there can be
actual physical  effects arising from what philosophers refer to as
{\em counterfactuals} -- that is, things that might have happened,
although they did not happened. (1994, 240)
\end{quotation}
In particular, Penrose's quotation relates to interaction-free
measurements (Elitzur and Vaidman, 1993) in which a location of a
supersensitive mine, which explodes if anything ``touches'' it, can be
found without an explosion. The counterfactual here is the explosion
which could have happen, but didn't. What allows such counterfactuals
without miracles is the {\it indeterminism} of the quantum theory
(with collapse).  In the non-collapse deterministic interpretation
such as the Many-Worlds Interpretation of quantum theory (Everett
1957) the explanation is different (and in my opinion is particularly
clear).  The counterfactuals are ``actual'' in other worlds (Deutsch 1997, 275).  Thus, in the situations considered by Penrose,
``things'' did happened in the physical universe (the union of all
worlds) and thus their effect on some other facts in the physical
universe is not so surprising (see Vaidman 1994).

The counterfactuals of the first type are certainly helpful: they
provide deeper explanation of many peculiar quantum phenomena. For
example, we can understand why there is an ``interaction-free''
measurement which can ascertain that in a certain location there is a
supersensitive mine, but there is no ``interaction-free'' measurement
ascertaining that in a certain location there is {\it no}
supersensitive mine: in the latter there is no counterfactual world
(such as the world with the explosion in the previous case) different
from the actual one.  However, quantum counterfactuals of the first
type cannot be brought to the general form (i) and they will not be
the main topic of this paper.

Quantum counterfactuals of the second type are  statements in the  form
(i) related to  a close quantum system.
 $\cal A$ defines which experiments are performed on this
system by an external observer and $\cal B$ is related to the results of
these experiments. The decision of the observer which experiments to
perform  is assumed to be
independent on the state of the quantum system under investigation. One can freely  change
everything outside the quantum system in question. This aspect
represents a  crucial difference between quantum counterfactuals and
the counterfactuals in the general
philosophical literature where  $\cal A$ is related to the whole world.

Most  examples of quantum counterfactuals discussed in the
literature are in the context of EPR-Bell type experiments, see Skyrms
(1982), Peres (1993), Mermin (1989) (which, however, does not use the
word counterfactual),  Bedford and Stapp (1995) who even present an
analysis of a Bell-type argument in the formal language of the Lewis
(1973) theory of counterfactuals, and Stapp (1997a) which followed by 
intensive polemic: Unruh (1997), Mermin (1997a,b), Finkelstein (1998),
Mashkevich (1998), and Stapp (1997b,c,d; 1998).
A typical
example is a consideration  of an array of
incompatible measurements on a composite system in  an entangled
 state.
 Various conclusions are derived from statements about the
results of these measurements. Since these measurements are
incompatible they cannot be all performed together, so it must be that
at least some of them were not actually performed.  This is why they
are called counterfactual statements.

Quantum counterfactuals are usually explicitly asymmetric in time. The
asymmetry is neither in $\cal A$ nor in $\cal B$; both are about the
{\it present} time.  The asymmetry is in the description of the actual
world.  The {\it past} but not the {\it future} of a system is given.

My purpose here is to avoid the asymmetry in time and to allow both
the past and the future of counterfactual worlds to be fixed. However,
it seems that  $\cal A$ changes
the future and therefore the future cannot be kept fixed. Indeed, the complete
description of a quantum system is given by its quantum state and the
choice of measurements, described by $\cal A$, changes the future
quantum state to be one of the eigenstate of the measured variable.
Therefore, we cannot hold fixed the quantum state of the system in the
future.

The way to overcome this difficulty  is not to
 use a quantum state as the  description of a physical  system.
For solving the current problem we can 
consider the quantum state only as a mathematical tool for calculating
the probabilities of the results of measurements, and not as a description
of the ``reality'' of a quantum system. 
Indeed, counterfactual statements are related to our experience which is connected
to a quantum system through results of experiments. Therefore, we can
define counterfactuals in terms of results of experiments without entering the
issue  the ``reality'' of a quantum system. The advantage
of this pragmatic approach  is that it is universal: it  fits all 
interpretations of quantum theory. Thus, my proposal for defining counterfactuals
in quantum theory is as follows:

\begin{quotation}
  { \bf (ii)} {\em If it were that measurement $\cal M'$ instead of
  measurement $\cal M$ has been performed on a system $S$, then it
  would be that the outcome of $\cal M'$ has property $\cal P$.  The
  results of all other measurements performed on the system $S$ are
  fixed.}
\end{quotation}
 $\cal M$ and $\cal M'$  consist, in general, of
measurements of several observables  performed at space-time points $P_i$. The
property $\cal P$ is a certain
relation between the results of these measurements  or a
probability for such relation to happen.

What makes my definition different and rigorous is the clarification
of what is fixed. Usually, this is not spelled out and it is tacitly
assumed that the quantum state of the system  prior to the times of the
space-time points $P_i$
is fixed.

In usual time-asymmetric situations, in which the past relative to
$P_i$ exists but the future does not, the counterfactuals according to my
definition are identical to those in the usual approach. Indeed, the
results of all measurements in the past define the quantum state
uniquely.  No controversies appear in such cases: the past of the
counterfactual worlds is fixed to be the past of the actual world.
The problems arise when there is some information about the future of
a system, sometimes, ``future'' only according to a particular Lorentz
frame.  (For systems consisting of spatially separated parts the
``past'' and ``future'' depends on the choice of the Lorentz frame.)
Following the principle that only the past is fixed and bringing together
``true'' counterfactuals from various Lorentz frames frequently leads to
paradoxes, see Hardy (1992a), Clifton et al. (1992), Stapp (1997a). In contrast, my definition (ii)
is unambiguous in such situations. It yields well defined statements
when  we are given results of measurements both in the past and in
the future of $P_i$, and in the cases when the space-time points $P_i$
are such that future and past cannot be unambiguously defined.

For a simple time-symmetric case in which $\cal M'$ describes a
single measurement of a variable $A$ performed between two complete
measurements which fix the states $|\Psi_1\rangle$ at $t_1$ and
$|\Psi_2\rangle$ at $t_2$, the definition (ii) becomes
\begin{quotation}
{ \bf (iii)}
{\em  If it were that a measurement of an observable $A$ has been performed at
  time $t$, $t_1<t<t_2$, then  the probability for
$A=a_i$ would be equal to $p_i$, provided  that the results of  measurements
performed on the system at times $t_1$ and $t_2$  are fixed.}
\end{quotation}
 The probabilities $p_i$ are given by the ABL formula (Aharonov et
 al. 1964; Aharonov and Vaidman 1991): 
\begin{equation}
  \label{ABL}
 {\rm Prob}(a_i) \equiv p_i = {{|\langle \Psi_2 | {\bf P}_{A=a_i} | \Psi_1 \rangle |^2}
\over{\sum_j|\langle \Psi_2 | {\bf P}_{A=a_j} | \Psi_1 \rangle |^2}} .
\end{equation}
The application of the time-symmetric formula (1) to counterfactual
situations led to a considerable controversy, see Albert et al. (1985),
Bub and Brown (1986), Sharp and Shanks
(1993),  Cohen (1995), Miller (1996), Vaidman (1998).  I believe, that
the time-symmetric definition (iii) provides a consistent way for
application of the ABL rule for counterfactual situations, thus
resolving the controversy, see more details in Vaidman (1996).

The definition (iii) is also helpful in analyzing  various attempts to prove
that {\it realistic} quantum theory leads to a contradiction with relativistic
causality. The  ``element of reality'' can be considered as an
example of a counterfactual (iii)  in the particular case of
probability 1 for a certain outcome. A time-symmetrized definition of element of
 reality is (Vaidman 1993a):
\begin{quotation}
{ \bf (iv)}  {\em If we can {\em infer} with certainty that the result of
  a measurement  at time $t$ of an observable $A$ is $a$, then, at time
   $t$, there exists an element of reality $A=a$.}
\end{quotation}
The word ``infer'' is neutral relative to past and future. The
inference about results at time $t$ is based on the results of
measurements on the system performed both before and after time $t$.

An important  feature of time-symmetric  elements of reality
(iv) of a pre- and  post-selected quantum system is that the
``product rule'' does not hold. The product rule means that  if $A=a$ and $B=b$ are elements
of reality, then $AB =ab$ is also an  element of reality.

A simple example of this kind is a system of two spin-${1\over 2}$
particles prepared at $t_1$ in a singlet state
\begin{equation}
|\Psi_1\rangle = {1\over {\sqrt 2}}(  |{{\uparrow}}\rangle_1
|{\downarrow}\rangle_2 -   |{\downarrow}\rangle_1
|{{\uparrow}}\rangle_2).
\end{equation}
At $t_2$ the particles are found in the
state
\begin{equation}
|\Psi_2\rangle = |{\uparrow}_x\rangle_1 |{\uparrow}_y\rangle_2  .
\end{equation}
 A  set of elements of reality for these particles at an
intermediate time $t$ is (use the ABL formula (1) to see this):
\begin{eqnarray} 
\{{\sigma_1}_y\} = -1 , \\
\{{\sigma_2}_x\} = -1 ,\\
\{{\sigma_1}_y {\sigma_2}_x\}=-1 .
\end{eqnarray}
where the notation $\{X\}$ signifies the outcome of a measurement of
$X$. Indeed, the product rule does not hold: $\{{\sigma_1}_y
{\sigma_2}_x\} \neq \{{\sigma_1}_y\} \{{\sigma_2}_x\}$. Note that a
measurement of the nonlocal variable in Eq. (6), the product of local
variables related to separated locations, is not disallowed due to
locality of physical interactions. This particular measurement can be
performed using local interactions only (Aharonov et al. 1986).

The failure of the product rule plays an important role in discussing
Lorentz invariance of a realistic quantum theory, especially, in the
light of recent proposals to prove the impossibility of a realistic
Lorentz invariant quantum theory which applied the product rule (Hardy
1992a; Clifton et al. 1992). See more discussion of this controversy
in Vaidman (1993b, 1997), Cohen and Hiley (1995, 1996).

It seems to me that the proposed definition for counterfactuals should
also help to resolve the recent controversy 
(Stapp 1997a,b,c,d, 1998; Unruh 1997; Mermin 1997a,b; Finkelstein 1998;
Mashkevich 1998) generated by 
the proposal of  Stapp
to prove  nonlocality of quantum
theory   using  counterfactual analysis of Hardy-type experiment
(Hardy 1992b). My definition resolves the vagueness in these
discussions,  pointed out by
Finkelstein (1998), about what is fixed in the counterfactual worlds.

I  claim that quantum theory does not support the second
locality condition of Stapp (his LOC2). Stapp considers two spatially
separated spin-${1\over 2}$ particles. In his example, a certain
counterfactual statement related to a particle on the right can be proved
given that certain action was performed before that on a spatially
separated particle on the left.  He then notes that in another Lorentz
frame the action on the particle on the left is performed {\it after}
the time to which the counterfactual statement is related.  Stapp
concludes that since an action in the future cannot influence the
past, the action on the left side can be replaced by some other action
without changing the truth of the counterfactual related to the
particle on the right.

The argument which led Stapp to his locality condition LOC2 does not go through if we adopt the definition
of counterfactuals (ii),  considering measurements on the
particle on the right while  keeping fixed the results of all other
measurements on our system (the system consisting ot the two
spin-${1\over 2}$ particles).  Then, the truth of the counterfactual
requires only the existence of a Lorentz frame in which the
measurements on the right side are after the measurement on the left,
and the consideration of the other Lorentz frames is irrelevant.

In Stapp's example we indeed have a situation in which an action in a
space-like separated region on the left side  changes the truth of 
a certain counterfactual statement about measurement on the right.
However, I do not see that the failure of LOC2 proves the ``nonlocal
character of quantum theory'' as the title and the spirit of Stapp's
paper suggest. In order to demonstrate the meaning of LOC2 let me
present another example where it fails.

Consider again  
 two spatially separated 
spin-${1\over 2}$ particles prepared in a singlet state (2). At time
$t$ a Stern-Gerlach experiment with the gradient of a magnetic field in the
positive $\hat z$ direction  is performed and the result
${\sigma_2}_z= \alpha$  is obtained.
 Now consider the following counterfactual statement:

\begin{quotation}{\bf CF}: {\em If it were that the measurement has been performed with the gradient pointing in
the negative $\hat z$ direction instead, then the same result ${\sigma_2}_z= \alpha$ would be
obtained.}
\end{quotation}
The truth of this statement depends on actions on particle 1 in a
space-like separated region. Indeed, if the measurement of ${\sigma_1}_z$
was performed, the CF is true, if no measurement is performed or, say
${\sigma_1}_x$ were measured, then the truth of CF does not
follow. (Note, that according to the Bohm-Bell hidden variable
interpretation (Bohm 1952, Bell 1987),  CF must be false in this case.)

Of course, since CF cannot be tested,  no contradiction with
relativistic causality can arise. Still, there is some nonlocality
 in this example. For me, the framework of the MWI yields  the
clearest picture of  this
nonlocality. By performing measurements
on particle 1 we split our world into two worlds creating a {\it
  mixture} of two worlds for particle 2. By different choices of
measurement on particle 1 we create different mixtures of 
worlds for particle 2. For example, the two worlds with definite ${\sigma_2}_z$
(for which  CF is true) or the two worlds with definite ${\sigma_2}_x$ (for
which CF does not follow). Although the worlds are different, the two
mixtures  are physically equivalent for particle 2 and therefore there was no nonlocal action
in the physical universe which incorporates all the worlds. The nonlocality is as follows: the 
 world (branch) in the MWI is a nonlocal entity which, in our case,
is defined by  properties in the location of the two particles. The choice of a
local measurement on particle 1 defines the set of worlds into which
the present world will be split. In this way an action on particle 1 
leads to various sets of possible properties related to particle 2.

I have to mention a property of definition (ii) which might be
considered as its weakness. The outcome of measurement $\cal M$
performed in the actual world plays no role in calculating the truth of
the counterfactual statement (except trivial cases in which $\cal P$ involves a
comparison between the outcome of $\cal M$ and $\cal M'$ as in the previous
example). It is assumed that properties of the outcome of $\cal M'$ are
independent on the outcome of $\cal M$. This is what standard quantum
theory tells us, but this is not true, in general, for hidden variable
theories: The outcome of $\cal M$ can yield certain information about
hidden variables, the information which might help to ascertain the
properties of the outcome of $\cal M'$. However, in the framework of
the hidden variables theories, the definition (ii) seems to be 
incomplete anyway. In order to avoid vagueness we must add a
statement about hidden variables, for example, by fixing hidden
variables in a counterfactual world to be equal to the hidden
variables in the actual world. Given this correction, the outcome of
$\cal M$ adds no information again. However, I do not want to adopt
this approach because it explicitly time-asymmetric: the hidden
variables are fixed only in the past. I do not know how to approach the
problem of time-symmetric hidden variables.

The proposed definitions of counterfactuals (ii) and (iii) are also
applicable for counterfactuals in classical physics. However, due to
the determinism of classical theory we cannot fix independently the
results of a complete set of measurements in the past and the results
of the complete set of measurements in the future.  Note that there
are certain limitations of this kind  in the quantum case too. For example, consider
a spin-$1\over 2$ particle with three consecutive measurements
${\sigma}_z(t_1)=1$, ${\sigma}_x(t)=1$ and ${\sigma}_z(t_2)=-1$,
$t_1<t<t_2$. Then a counterfactual statement ``If at time $t$ a
measurement of ${\sigma}_z$ were performed instead, the result would
be ${\sigma}_z(t)=1$'' is neither true, nor false, but meaningless
because the results of measurements ${\sigma}_z(t_1)=1$, and
${\sigma}_z(t_2)=-1$ are impossible when ${\sigma}_z$, instead of
${\sigma}_x$, is measured at time $t$.  Nevertheless, such constrains
are not strong and they leave a room for numerous nontrivial
counterfactuals.

In  classical physics the
counterfactuals (ii) have even more serious problem. 
$\cal M'$ consist of measurements of some observables. We can make a one to
one correspondence between ``the outcome of a measurement of an
observable $O$ is $o_i$'' and ``the value of $O$ is $o_i$''. The
latter is independent of whether the measurement of $O$ has been
performed or not and, therefore, statements which are formally
counterfactual about results of possible measurements can be replaced
by ``factual'' (unconditional) statements about values of
corresponding observables.  In contrast, in standard quantum theory,
observables, in general, do not have definite values and therefore we
cannot always reduce the above counterfactual statements to
``factual'' statements.

I do not expect that everybody will agree with my proposals for
resolving the controversies discussed above. I hope only that the main
result of this work will not be controversial: a consistent definition
of counterfactuals in quantum theory, a definition that is equivalent
to the standard approach for the time-asymmetric cases in which only
the past of the system is given, but which is applicable to the
time-symmetric situation (such as pre- and post-selected systems) -- a
definition which is a useful tool for the analysis of many current
problems.

  The research was
supported in part by grant 614/95 of the Israel Science Foundation.

\vskip .8cm

 \centerline{\bf REFERENCES}
\vskip .15cm
\footnotesize

\vskip .13cm \noindent 
 Aharonov, Y., Albert, D.,  and Vaidman, L. (1986),
 ``Measurement Process in Relativistic Quantum Theory'',
{\em Physical  Review  D 34}: 1805-1813.

\vskip .13cm \noindent 
Aharonov, Y.,  Bergmann,  P.G., and  Lebowitz, J.L. (1964),
``Time Symmetry in the Quantum Process of Measurement'',
 {\em Physical Review 134B}: 1410-1416.

\vskip .13cm \noindent 
Aharonov, Y. and Vaidman, L. (1991),
``Complete Description of a Quantum System at a Given Time'',
{\em Journal of  Physics  A 24}: 2315-2328.

\vskip .13cm \noindent 
Albert, D.,  Aharonov, Y. and  D'Amato, S. (1985),
``A Curious New Statistical Prediction of Quantum Theory''
{\em Physical Review Letters 54}: 5-8.

\vskip .13cm \noindent 
Bedford, D. and Stapp, H. P. (1995),
``Bell's Theorem in an Indeterministic Universe'',
{\em Synthese 102}: 139-164.

\vskip .32 cm \noindent
 Bell, J. S. (1987), ``Quantum  Mechanics for Cosmologists'', in Bell,
 J. S., {\em Speakable and Unspeakable in Quantum Mechanics}.
Cambridge: Cambridge University Press, pp. 117-138.

\vskip .13cm \noindent 
Bennett, J. (1984),
``Counterfactuals and Temporal Direction'',
{\em Philosophical Review 93}: 57-91.

\vskip .13 cm \noindent
 Bohm, D. (1952), ``A Suggested Interpretation of the Quantum Theory in
 Terms of `Hidden' Variables I and II,'' {\em Physical Review 85}: 97-117.

\vskip .13cm \noindent 
 Bub, J. and Brown, H. (1986), 
``Curious Properties of Quantum Ensembles which have been Both
  Preselected and Post--Selected'',
{\em Physical Review Letters 56}: 2337-2340.

\vskip .13cm \noindent 
Clifton, R., Pagonis, C., and Pitowsky, I.  (1992 ), 
``Relativity, quantum Mechanics and  EPR''
 {\em Philosophy of Science Association 1992}, Volume 1: 114-128.

\vskip .13cm \noindent 
Cohen,  O. (1995), 
``Pre- and Post-Selected Quantum Systems, Counterfactual Measurements,
and Consistent Histories'',
  {\em Physical Review  A 51}: 4373-4380. 
 
\vskip .13cm \noindent 
 Cohen, O. and Hiley, B.J. (1995),
  ``Reexamining the Assumptions that Elements of Reality can be
  Lorentz Invariant'', {\em Physical Review A 52}: 76-81.

\vskip .13cm \noindent 
  Cohen, O. and Hiley, B.J. (1996), 
``Elements of Reality, Lorentz Invariance and the Product Rule'',
 {\em Foundations of Physics 26}: 1-15.

\vskip .13cm \noindent 
Deutsch, D. (1997),
 {\em The Fabric of Reality},
New York: Allen Lane The Penguin Press. 

\vskip .13cm \noindent 
Elitzur, A. and Vaidman, L. (1993),
``Interaction-Free Quantum Measurements,''
 {\it Foundation of Physics  23}: 987-997.

\vskip .13cm \noindent 
Everett, H. (1957),
 `` `Relative State' Formulation of Quantum Mechanics'',
{\em Review of Modern Physics 29}: 454-462.

\vskip .13cm \noindent 
Finkelstein, J. (1998),
 ``Yet Another Comment on  `Nonlocal Character of Quantum Theory'',
preprint  quant-ph/9801011.

\vskip .13cm \noindent 
Hardy, L. (1992a)
``Quantum Mechanics, Local Realistic Theories, and Lorentz-Invariant Realistic Theories'',
{\em Physical Review Letters 68}: 2981-2984.

\vskip .13cm \noindent 
Hardy, L. (1992b)
``A Quantum Optical Experiment to Test Local Realism'',
{\em Physics  Letters A 167}: 17-23.

\vskip .13cm \noindent 
Lewis, D. (1973),
{\em Counterfactuals}.
Oxford: Blackwell Press.

\vskip .13cm \noindent 
 Lewis, D. (1986),
 ``Counterfactual Dependence and
  Time's Arrow'' reprinted from {\em Nous 13}: 455-476 (1979), 
 in Lewis, D. {\em  Philosophical Papers Vol.II}, Oxford: Oxford 
University Press,
  pp.32-52.

\vskip .13cm \noindent 
Mashkevich V. S. (1998),
``On Stapp-Unruh-Mermin Discussion on Quantum Nonlocality: Quantum
Jumps and Relativity'', preprint quant-ph/9801032.

\vskip .13cm \noindent 
Mermin, N.D. (1989),
``Can You Help Your Team Tonight by Watching on TV? More Experimental
Metaphysics from Einstein, Podolsky, and Rosen'',
in J.T. Cushing and E. McMullin (eds.) {\em Philosophical Consequences
  of Quantum Theory: Reflections on Bell's Theorem}. Notre Dame:
University of Notre Dame Press, pp.38-59.

\vskip .13cm \noindent 
Mermin, N.D. (1990),
``Quantum Mysteries Revisited'',
{\em American Journal of Physics 58}: 731-734.

\vskip .13cm \noindent 
Mermin, N.D. (1997a),
  ``Nonlocal Character of Quantum Theory?'',
preprint  quant-ph/9711052.

\vskip .13cm \noindent 
Mermin, N.D. (1997b),
``Nonlocality and Bohr's Reply to EPR'', preprint  quant-ph/9712003. 

\vskip .13cm \noindent 
Miller, D. J. (1996),
``Realism and Time Symmetry in Quantum Mechanics'',
{\em Physical Letters A 222}: 31-36.

\vskip .13cm \noindent 
Penrose, R. (1994),
{\em Shadows of the Mind}.
 Oxford: Oxford University Press.

\vskip .13cm \noindent 
Sharp, W.D. and Shanks, N. (1993),
``The Rise and Fall of Time-Symmetrized Quantum Mechanics'',
{\em Philosophy of Science 60}: 488-499.

\vskip .13cm \noindent 
 Skyrms, B. (1982),
 ``Counterfactual Definetness and Local
 Causation'', {\em Philosophy of Science 49}: 43-50.

\vskip .13cm \noindent 
 Stapp, H. P. (1997a),  ``Nonlocal Character of Quantum Theory'',
{\em American  Journal of  Physics  65}: 300-304.

\vskip .13cm \noindent 
 Stapp, H. P. (1997b),
``Mermin's Suggestion and the Nature of Bohr's
    Action-at-a-Distance Influence'', preprint quant-ph/9711060.

\vskip .13cm \noindent 
 Stapp, H. P. (1997c),
``Nonlocality and Bohr's reply to EPR'', preprint  quant-ph/9712036. 

\vskip .13cm \noindent 
 Stapp, H. P. (1997d),
``Reply to Unruh'', preprint quant-ph/9712043.

\vskip .13cm \noindent 
 Stapp, H. P. (1998),
``Comments on  Unruh's Paper'', preprint quant-ph/9801056.

\vskip .13cm \noindent 
 Vaidman, L. (1993a),
``Lorentz-Invariant `Elements of Reality' and the Joint
Measurability of Commuting Observables'',
{\em Physical Review Letters 70}: 3369-3372.

\vskip .13cm \noindent 
Vaidman, L. (1993b),
`` `Elements of Reality' and the Failure of the Product Rule''
  in P.J. Lahti, P.~Bush, and P. Mittelstaedt (eds.), {\em Symposium on the Foundations  of Modern Physics}. New Jersey: World
Scientific, pp. 406-417.

\vskip .13cm \noindent 
Vaidman, L. (1994),
``On the Paradoxical Aspects of New Quantum Experiments'',
 {\it Philosophy of Science Association 1994}, pp. 211-217.

\vskip .13cm \noindent 
Vaidman, L. (1996), ``Defending Time-Symmetrized Quantum Theory'', preprint
quant-ph/9609007.

\vskip .13cm \noindent 
Vaidman, L. (1997),
 ``The Analysis of Hardy's Experiment Revisited'', preprint  quant-ph/9703018.

\vskip .13cm \noindent 
Vaidman, L. (1998),
``On the Validity of the Aharonov-Bergmann-Lebowitz Rule'',
 {\em Physical Review  A}, to be published, November 1998,  quant-ph/9703001.

\vskip .13cm \noindent 
Unruh, W. G. (1997),
``Is Quantum Mechanics Non-Local?'', 
      preprint  quant-ph/9710032.

\end{document}